\documentclass[prl,twocolumn,letterpaper,showpacs,amsmath,amssymb,floatfix]{revtex4}
\usepackage[dvips]{graphicx}
\usepackage{amssymb}
\begin{document}

\newcommand{\ket}[1]{|#1\rangle}
\newcommand{\bra}[1]{\langle #1|}
\newcommand{\bracket}[2]{\langle #1|#2\rangle}
\newcommand{\ketbra}[1]{|#1\rangle\langle #1|}
\newcommand{\average}[1]{\langle #1\rangle}
\newtheorem{theorem}{Theorem}

\title{Equiangular Spherical Codes in Quantum Cryptography}
\author{Joseph M. Renes}
\affiliation{Department of Physics and Astronomy, University of New Mexico,\\
Albuquerque, New Mexico 87131--1156, USA\\
\texttt{renes@phys.unm.edu}}

\begin{abstract}
Quantum key distribution protocols based on equiangular 
spherical codes are introduced and their behavior under the intercept/resend
attack investigated. Such protocols offer a greater range of secure noise tolerance and speed options than
protocols based on their cousins, the mutually-unbiased bases, 
while also enabling the determination of the channel noise rate 
without the need to sacrifice key bits. For fixed number of signal states 
in a given dimension, the spherical code protocols offer Alice and Bob more
noise tolerance at the price of slower key generation rates. 
\end{abstract}


\maketitle

\noindent The possibility of secure key distribution using quantum
states is by now a well established feature of quantum information
theory.  In the original 1984 proposal of Bennett and Brassard
(BB84)~\cite{bb84}, four states of a spin-1/2 system, the eigenstates
of $\sigma_z$ and of $\sigma_x$, are used as signals by the sender
Alice. These states are naturally partitioned into two orthonormal
bases from which the receiver Bob chooses one at random to measure the
signal. Because the bases are \emph{unbiased}---i.e., the overlap
between vectors from distinct bases is always the same, equal to $1/2$
for qubits---Bob learns nothing when his measurement doesn't correspond
to Alice's preparation, but everything when it does. The
nonorthogonality of all the states allows Alice and Bob to detect
eavesdropping by an adversary Eve, so the states form an
unconditionally secure cryptographic protocol~\cite{mayers98,biham00,lochau99,shorpreskill00}.

One more unbiased basis, the eigenvectors of $\sigma_y$, can be added
to the BB84 set, forming a six-state protocol~\cite{bruss98}.
Unbiased bases can be found in higher dimensions as
well~\cite{woottersfields89}, and the key distribution protocol has
been extended to such cases, with increasing dimension leading to
improved security~\cite{cbkg02}. 

Here we introduce a new ensemble of signals, equiangular spherical codes, 
and show that key distribution protcols based on them offer potential advantages 
over the unbiased bases. Our security analysis is based on the 
eavesdropper's use of the intercept/resend attack and, as such, doesn't provide
firm proofs of security, but rather yields an insight into the workings and features 
of the protocol. The first feature is simplicity. Such protocols need not sacrifice potential
key letters in order to establish the amount of information Eve has learned
about the whole key string. The success rate (sift rate) of the protocol itself furnishes this
information. Second, equiangular spherical codes offer a greater range of 
protocol security and key generation speed options for fixed dimension.
Finally, for fixed number of signal states in a given dimension, these protocols
offer a higher noise threshold for secure operation but lower key generation rates than do the mutually-unbiased bases.



Recall the general setting of quantum key distribution. Two parties,
Alice and Bob, wish to make use of an authenticated public classical broadcast
channel and an insecure quantum channel controlled by an adversary 
Eve to establish a secret key for the purposes of encrypting and sharing
other data. They start by using the classical channel to fix a signal ensemble and a
measurement for the quantum channel. Alice sends states drawn from the
signal ensemble through the quantum channel to Bob, who performs the
chosen measurement (in the case of signaling states drawn from
mutually unbiased bases, the several measurement bases Bob chooses from
for his measurement are here amalgamated into a single generalized
measurement). Eve is free to exploit knowledge of the protocol 
and her control of the quantum channel
to mount an attack on their protocol; she can in principle subject the signal states to
any physical interaction that she wishes.
Effectively, this process produces a sequence of samples from a
certain tripartite probability distribution shared between the three parties. 
Alice and Bob then proceed to ``distill'' the key by 
communicating information based on their individual sequences over the classical channel.
Their goal is to exploit the quantum nature of the channel to make Eve's eavesdropping
ineffective.

The relevant probability distribution is the joint probability
$p(a_i,b_j,e_k)$ of Alice's signal, Bob's measurement result, and the
result of whatever measurement Eve performs in the course of eavesdropping.
Repeated use of the protocol yields a sequence of samples drawn from
this distribution.  Alice and Bob, however, must establish which
distribution they are sampling from, as it depends on Eve's attack.
We imagine Eve has some physical setup which can give rise to many
different distributions as she changes the strength of her interference
with the channel. Given an assumption of the type of attack, Alice and
Bob determine the extent of Eve's interference by making public and
comparing a fraction of the Alice's signals and Bob's measurement
results. In this way they estimate the error rate of the channel, and
together with an assumption of the attack, determine the distribution $p$.
From the remaining samples, which are supposed to be an asympotically 
large number $M$, say, they can distill a key of
length $MR$ in accordance with the following bounds:
\begin{equation}
\label{eq:keyratebound}
I_E\leq R\leq I(A\!:\!B|E)\;,
\end{equation}
where $I(X\!:\!Y)=H(X)+H(Y)-H(XY)$ is the mutual information of $X$ and
$Y$, $H(\cdot)$ being the Shannon entropy, and
$I_E=I(A\!:\!B)-\min\{I(A\!:\!E),I(B\!:\!E)\}$.  The lower bound
obtains when the key is distilled using one-way
communication~\cite{ck78}; to progress beyond this requires a technique
called \emph{advantage distillation}, though this is of limited
efficiency~\cite{maurer93,gisinwolf99}.

These bounds provide a method of investigating the cryptographic
usefulness of a protocol. Given a signal ensemble, Bob's
measurement, and an assumption about the nature of Eve's attack, the
probability distribution can be calculated, and the key rate bounds
determined.  In this way the security of the protocol against this
attack is established.  To say that a protocol is unconditionally
secure is to demonstrate its security against all possible attacks.

The focus now turns to Alice's signal ensemble and Bob's measurement.
An intuitively appealing ensemble is a \emph{spherical code}, a
complex-vector-space version of points on a sphere whose minimal
pairwise distance is maximal.  The complex version, called the
\emph{Grassmann packing problem}, asks for a set of unit vectors in
$\mathbb{C}^d$ whose \emph{maximal\/} pairwise overlap is
\emph{minimal}~\cite{Strohmer03}. When all these pairwise overlaps are
equal, this \emph{equiangular\/} spherical code is called a
\emph{Grassmann frame}; i.e., a set
$\mathcal{C}=\{\ket{\phi_k}\in\mathbb{C}^d\}_{k=1}^n$ for $n\!\geq\! d$
is a Grassmann frame if
\begin{equation}
\label{eq:sc}
|\bracket{\phi_j}{\phi_k}|^2=\frac{n\!-\!d}{d(n\!-\!1)}\qquad \forall\,\,
j\!\neq\! k\;.
\end{equation}

Grassmann frames also arise as the solution to the ``minimum energy
problem.''  For a set of unit vectors $\mathcal{C}$, call
$V_t(\mathcal{C})=\sum_{j,k}|\bracket{\phi_j}{\phi_k}|^{2t}$ the $t$-th order
``potential energy'' of the set of the vectors~\cite{benedettofickus03}.  
The minimum energy
problem is to find $\mathcal{C}$ having $n\!\geq\! d$ elements such
that $V_1=n^2/d$ and $V_2$ is minimized.  Note that $n^2/d$ is the
global minimum of $V_1$.
This follows from considering the (at most) $d$ nonzero (real)
eigenvalues $\gamma_j$ of the Gram matrix
$G_{jk}\!=\!\bracket{\phi_j}{\phi_k}$. Clearly $\sum_k \gamma_k\!=\!n$
and $\sum_k \gamma_k^2\!=\!V_1(\mathcal{C})$.  These being the
equations for a plane and a sphere, the minimum of $V_1$ occurs if and
only if all the $\gamma_k$ are equal to $n/d$, whence $V_1$ is bounded
below by $n^2/d$.  Thus what is sought is the set of vectors with the
minimum $V_2$ energy, given minimum $V_1$ energy.

To find a lower bound for the minimum of $V_2$, let
$\lambda_{jk}=|\bracket{\phi_j}{\phi_k}|^2$, and employ the same method
again.  We have immediately that $\sum_{j\neq
k}\lambda_{jk}=V_1-n=n(n-d)/d$ and $\sum_{j\neq
k}\lambda_{jk}^2=V_2-n$, whence the minimum of $V_2$ over all sets
minimizing $V_1$ is bounded below by making all the $\lambda_{jk}$ the
same and given by Eq.~(\ref{eq:sc}).  When this lower bound is
achieved, i.e $V_2=n^2(n-2d+d^2)/(n-1)$, the result is a Grassman frame.

The existence of Grassmann frames isn't established for arbitrary $n$ and
$d$, though some general statements can be made~\cite{rbksc03}.  They always exist for
$n=d+1$ (a regular simplex), but never when $n>d^{\,2}$.  For $n\leq
d^{\,2}$, when a Grassman frame exists, it is a spherical code, but for
$n>d^{\,2}$, spherical codes aren't equiangular.

By minimizing $V_1$, Grassmann frames automatically form measurement
POVMs, which can be used by Bob to detect Alice's signal. This is true
because $S=\sum_k \ket{\phi_k}\bra{\phi_k}=(n/d)I$, so that a POVM can
be constructed from the subnormalized projectors
$(d/n)\ketbra{\phi_k}$.  To see this, fix an orthonormal basis
$\{\ket{e_k}\}$ and consider the matrix $T_{jk}=\bracket{e_j}{\phi_k}$.
The Gram matrix can be written as $G_{jk}=(T^\dagger T)_{jk}$, while
$S_{jk}=(TT^\dagger)_{jk}$, so both have the same eigenvalues.  When
$V_1$ is minimized, these $d$ eigenvalues are all $n/d$, implying that
the vectors form a resolution of the identity.

By using the same ensemble as Alice, Bob's measurement confirms the signal she sent
with probability $d/n$. Bob may also choose a measurement which attempts to 
repudiate some of Alice's signals, in a manner entirely similar to 
unambiguous state discrimination~\cite{chefles98}. Because the number of states is larger than the dimension
of the space on which they are supported, it is impossible to unambiguously determine
the signal Alice sent with nonzero probability. However, for any ensemble,
a \emph{parital} determination may be made, a scheme which works as
follows. First, partition the signal ensemble into the set of all subsets of size $b$. Then, for
each subset, find the projector orthogonal to the span of the vectors in the subset. 
This procedure yields a measurement operator for each of the subsets. 
In order to find an orthogonal projector to the span, $b$ must be restricted such
that none of the subsets spans more than $d-1$ dimensions. 

For general ensembles, the operators constructed by this procedure do not quite form a measurement; some additional
``failure'' outcome is required to make the entire set form a resolution of
identity. Interestingly, this additional outcome appears not to be required when the
signal ensemble is an equiangular spherical code. Numerical constructions starting with
ESCs in modest dimensions always yields a proper POVM in which no ``failure'' outcome
is necessary. Though both types of measurements deserve further study, we
shall specialize to the case of confirmation protocols; repudiation protocols are 
developed for qubits in~\cite{renes04}.

Confirmation protocols are appealing because the ESC ensembles are the sets that are ``least
classical'' in the following sense~\cite{fuchssasaki03a}.  Consider
using these quantum states as signals on a \emph{classical} channel as
follows.  Instead of sending the quantum state, Alice performs the
associated measurement and communicates the result to Bob using a
classical channel.  Bob then prepares the associated quantum state at
his end.  The fidelity of Bob's reconstruction with the input state,
averaged over inputs and measurement results, measures how well the
classical channel can be used to transmit quantum information.  This
fidelity is $dV_2/n^2$, so among all ensembles which themselves 
form POVMs, Grassmann frames are hardest to transmit
``cheaply'' in this way.  Eavesdropping on the communication between
Alice and Bob makes the channel more classical---Eve is essentially
trying to copy the signal---so one might expect that Grassmann frames
are useful in foiling the eavesdropper.

Before examining their resistance to eavesdropping, we remark on how
Alice and Bob can concretely use the equiangular spherical codes to accomplish secure 
key distribution. Though a protocol satisfying the lower bound of equation~\ref{eq:keyratebound} 
is guaranteed asymptotically, it may not be feasible in practice. 
Consider first the case of a noiseless quantum channel, i.e. no eavesdropping.
In a length-$N$ string of samples Alice and Bob will agree with probability $d/n$. 
When labelling the states $0$ to $n-1$, 
Bob's string $b$ is simply Alice's string $a$ plus a string $\delta$ having a 
fraction $(n\!-\!d)/n$ of non-zero elements. Alice can select a classical
error-correcting code $\mathcal{C}$ which can correct these errors,
choose a codeword $c$ randomly, and send $a+c$ to Bob. He then
simply subtracts this from his string to obtain $c+\delta$, from which
he uses the error-correcting property to determine the codeword $c$. 
From the Shannon noisy-channel coding theorem, there are roughly 
$N I(A\!:\!\!B)$ codewords, in
accordance with the lower bound. 

When Eve has no information about the quantum signals, the communicated
string $a+c$ tells her nothing about $c$, since it's effectively encrypted by $a$.
Should Eve have some information about $a$, gleaned from her tampering with the 
quantum channel, Alice and Bob may proceed as before to establish $c$, and 
then use a privacy amplification procedure to shorten this string and remove 
any information Eve has about it. 

As formulated, this protocol is not 
particularly robust; Alice and Bob can do better by first
announcing some of the signals not received, a procedure analagous to 
sifting in the case of mutually unbiased bases. Upon receipt of each
signal, Bob publicly broadcasts $m$ outcomes he did not obtain. If
Alice's signal is among these, they throw it away and proceed to the next. 
For the signals which pass the test Alice and Bob relabel the remaining
states in order from $0$ to $n-m-1$ and follow the above procedure. 
This occurs with 
probability $m/(n-1)$ as Bob could send any of ${n-1}\choose{m}$ outcomes 
and ${n-2}\choose{m-1}$ of these contain Alice's signal.
The protocol itself succeeds with probability
\begin{equation}
\label{eq:sucnonoise}
p_{\rm sift}=\frac{n(n\!-\!1)-m(n\!-\!d)}{n(n\!-\!1)}=\frac{s}{n(n\!-\!1)},
\end{equation}
where we have implicitly defined the constant $s$. Meanwhile, the key rate when using
an $n$-word equiangular spherical code in $d$ dimensions excluding $m$
outcomes is given by
\begin{eqnarray}
R&=&\log[n\!-\!m]+\frac{d(n\!-\!1)}{s}\log[d(n\!-\!1)]\nonumber\\
&&+\left(1-\frac{d(n\!-\!1)}{s}\right)\log[n\!-\!d]-\log s.
\end{eqnarray}

The real question is how well the protocol tolerates attempted eavesdropping.
To begin formulating an answer it is useful to consider the intercept/resend 
attack. Because it is straightforward to analyze, it offers immediate insight
into the usefulness of these protocols. A full treatment of the problem could
involve Alice and Bob using quantum error-correcting codes to turn
a noisy quantum channel into a smaller, noiseless channel. However, this
is far too complicated for the first step in this analysis. 

In the intercept/resend attack we assume that Eve fixes a fraction $q$ of 
signals to intercept. She measures those so chosen using the same equiangular
spherical code as does Bob, resending him the output of this process. Eve simply
guesses Alice's signal to be her outcome, unless it is excluded by Bob's 
announcement. In this case she may still guess, but retains the information that
she was forced to do so. 

By delineating the various cases, it is uncomplicated to arive at the relevant quantities.
First, the sift rate of the protocol depends on $q$ as in the following expression.
Letting $t=s(n-1)-qm(n-d)(d-1)$, the sift rate is now
\begin{equation}
p_{\rm sift}=t/n(n\!-\!1)^2.
\end{equation}
Alice and Bob's joint probability distribution is determined by the agreeement 
probability 
\begin{equation}
p_{a\!=\!b}=(n\!-\!1)(d(n\!-\!1)\!-\!q(n\!-\!d)(d\!-\!1))/t.
\end{equation}

Since Alice and Bob use the same ESC ensemble, Eve's joint probability with Alice 
is the same as with Bob. In order to account for the cases in which Eve measures the signal but
this outcome is later excluded by the protocol, we may append an event to her probability
distribution, denoted by \texttt{?}. Now she has $n-m+1$ total outcomes, and the \texttt{?}
outcome functions as a guess as to the key letter in the cases it occurs. 
The joint probability of such an exclusion and the particular signal $j$ is plainly the same for all $j$, 
and together with the probability of agreement between Alice and Eve, 
these quantities fully describe the overall distribution:
\begin{eqnarray}
p_{a=e}&=&q d (n\!-\!1) s/nt,\\
p_\texttt{?}&=&1-q s^2/nt.
\end{eqnarray}

These quantities enable us to compute the lower bound on the optimal 
key rate in equation~\ref{eq:keyratebound}. Then we may determine 
the value of $q$ such that $R=0$, and from this the maximum tolerable
error rate for a given $n,m,d$ combination. Finally, we can convert this error
rate into the corresponding channel depolarizing rate for the purposes of
comparison with unbiased bases protocols. Considering the error rate alone is 
misleading since for a noiseless channel Alice and Bob expect \emph{no}
errors, post sifting, when using unbiased bases, but do expect some errors when using the 
spherical codes. Better protocols remain secure at higher channel noise rates, and 
we pick the depolarizing channel for simplicity. For the spherical codes
the depolarizing rate $r$ is related to the error probability $p_{\rm e}=1-p_{a\!=\!b}$
via the expression
\begin{eqnarray}
r&=&\frac{s}{m(d\!-\!1)}\nonumber\\
&&-\frac{n(n\!-\!1)(n\!-\!m\!-\!1)}{m(d\!-\!1)(n\!-\!1+m(p_{\rm e}\!-\!1))}
\end{eqnarray}

For protocols using unbiased bases these expressions are much simpler;
in fact it is easier to work with the mutual information expressions themselves.
For a given probability of error between Alice and Bob, their shared information
is simply the full amount less the corruption caused by the error:
\begin{equation}
I(A\!:\!B)=\log d+p_{\rm e}\log p_{\rm e}+(1\!-p_{\rm e})\log(1-p_{\rm e})/(d-1)
\end{equation}
Eve gains information only if she measures in the correct basis, and the basis
announcement step prevents her from incorrectly guessing Alice's signal. Thus
she gains the full information $\log d$ if she manages to measure in the correct basis.
Letting $k$ be the number of bases, this makes Eve's information $q/k\log d$. 
The probability of error is $(k\!-\!1)(d\!-\!1)/kd$,
so in terms of error probability, Eve's mutual information is
\begin{equation}
I(A\!:\!E)=\frac{d}{(d\!-\!1)(k\!-\!1)}\,p_{\rm e}\log d
\end{equation}
The sift rate of the protocol is always the probability $1/k$ for Bob to measure in the
same basis Alice prepared the state, regardless of Eve's interference. Finally, the
equivalent depolarizing rate $r$ for a given error rate $p_{\rm e}$ is simply $r=p_{\rm e}d/(d\!-\!1)$.

Now we can compare the various aspects of both protocols. 
The first difference comes from the sift rate. When using
unbiased bases, the protocol will fail with probability $(k\!-\!1)/k$ no matter how noisy the
channel. To determine the error rate, Alice and Bob must announce some of their 
created key letters so as to compare how often they agree. Obviously these letters
cannot be used in the key so they are sacrificed. However, for equiangular
spherical codes the sift rate depends on the intercept rate. 
Successful execution of the protocol relies on the correlation between Bob's outcome
and Alice's signal, thus the probability of successful sifting decreases as the noise increases.
Further, Eve can't replace the signals with any others to increase the sift rate and
attempt to mask her intervention on other signals, for she would need to know Alice's 
signal. \emph{Thus, there is no need to sacrfice key letters in order to estimate Eve's
information.} 

The second immediate difference is the increased number of possible protocols using
the equiangular spherical codes. With unbiased bases Alice and Bob have the choice of 
using anywhere from 2 to (possibly) $d\!+\!1$ bases. 
Spherical codes offer more possibilities, as $n$ may range from $d\!+\!1$ to $d^2$ and $m$
from 0 to $n\!-\!2$. This offers two advantages. The first is speed. 
In particular there are protocols involving fewer than $2d$ signals,
the minimum for the unbiased bases. 
This translates into higher absolute key generation rates,
the key generation rate times the sift rate. 
Suppose Alice and Bob use $n=\alpha d$ spherical code states. Then the maximum key generation
rate becomes $\log(d)/\alpha+(\alpha\!-\!1)/\alpha\log((\alpha\!-\!1)/\alpha)$ for large $d$, in contrast to 
at best $\log(d)/2$ for unbiased bases. By choosing $\alpha<2$ Alice and Bob can 
find spherical codes which are faster than two unbiased bases. For example, consider 
35 spherical code states in 25 dimensions. The key rate is roughly 3.4 bits per signal
as compared to 2.3 for two unbiased bases.

The second advantage is security. 
Increasing $m$ is similar to increasing the number of unbiased bases in that both lead to 
decreased key rate and increased maximum tolerable noise. For maximum security
Bob may elect to announce all but two of the outcomes he didn't obtain, whereas for maximum
speed Bob would choose $m=0$. Figure 1 shows the maximum tolerable noise rate
for two values of $m$, as a function of number of signal states in ten dimensions. The 
corresponding maximum error rate when using various numbers of unbiased bases is included
for comparison. The case of $m=n-2$ always yields an improved secure noise threshold over unbiased bases. 
\begin{figure}
\label{fig:maxnoiserate}
\begin{center}
\includegraphics[scale=.8]{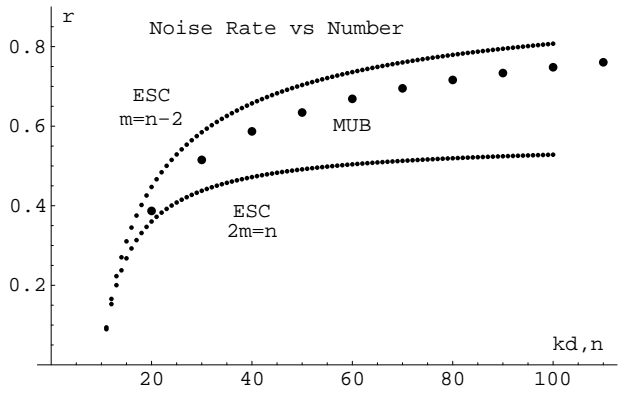}
\caption{Maximum tolerable noise rate in the depolarizing channel as a function
of number of elements in the equiangular spherical code ($n$) and mutually-unbiased 
bases ($kd$) ensembles, each in $d=10$ dimensions. 
Two ESC protocols are shown, above and below the 
MUB protocol. The upper curve corresponds to the maximal security case
$m=n-2$ in which Bob excludes all but two outcomes, 
while the lower corresponds to $m=n/2$, exclusion of half the outcomes. 
The flexibility in number of elements $n$ and number of ourcomes publicly excluded
$m$ provides an increased level of security over the unbiased bases.}
\end{center}
\end{figure}

These two advantages hold when considering \emph{all} possible ESC protocols in 
a fixed dimension and so are appropriate if only the dimension is constrained by
the particular hardware Alice and Bob wish to use. 
However, a direct comparison of the two protocols involves a tradeoff between speed and security
for a \emph{fixed} number of states in a given dimension. When using spherical code protocols, 
Alice and Bob are
free to choose $m$ to match the perceived error rate in the channel, so for each style of 
signal ensemble, fixing $n$ and $d$ specifies a concrete physical setup with similar resources.
For each protocol we may consider the pair consisting of maximum key generation rate and 
the maximum tolerable error rate. Plotting the pairs for protocols having $n$ signals in various
dimensions we can determine the tradeoff for the two protocols. 
\begin{figure}[h]
\label{fig:securecomp}
\begin{center}
\includegraphics[scale=.8]{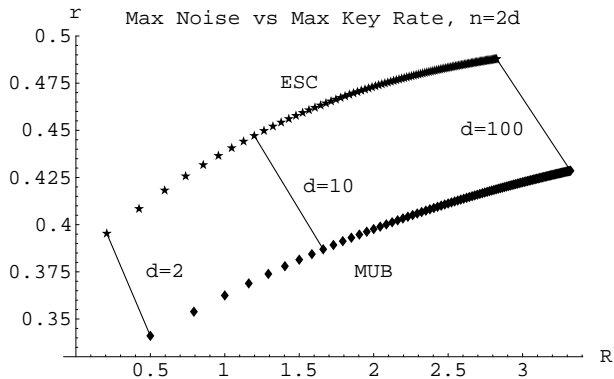}
\caption{Speed/security pairs for ESC and MUB protocols using $n=2d$ signals. Each point
represents the maximum key generation rate (horizontal axis) and maximum tolerable noise
rate for the depolarizing channel (vertical axis) for that particular protocol. The three lines
connect corresponding protocols between the two different signal ensembles and show 
that while the unbiased bases are faster for fixed $n$ and $d$, the spherical codes are more
robust.}
\end{center}
\end{figure}
Figure 2 reveals that for
$n=2d$ signals, the spherical code protocols offer more security (larger vertical values) but
at the price of slower key generation (smaller horizontal values). 

Use of spherical codes has been applied here to a specific model of quantum 
key distribution. However, they are immediately applicable to two variants of the
``prepare \& measure'' protocol discussed here. First is the coherent version
of such protocols in which Alice prepares a bipartite state, ostensibly entangled,
and sends half to Bob. Each party then measures his or her half, returning the 
protocol to the original picture. 
This method allows us to bring not only classical but quantum
information processing tools to bear on the problem of security. 
Equiangular spherical codes fit nicely into this framework, as they can
always be realized from maximally entangled states.  Thus they start on
the same footing as unbiased bases, for which this is also true. To
demonstrate this, consider a spherical code
$\mathcal{C}=\{\ket{\phi_k}\}$ and a ``conjugate'' code
$\mathcal{C}^*=\{\ket{\phi_k^*}\}$ formed by complex conjugating each
code state in the standard basis.  Then it is a simple matter to show
that $\ket{\Phi}=(\sqrt{d}/n)\sum_k\ket{\phi_k}\ket{\phi^*_k}$ is
maximally entangled.  Thus if Alice prepares this state and sends the
second half to Bob, they can realize the ``prepare-and-measure'' scheme
by measurement.

Second, the full array of $d^2$ equiangular states is a tomographically-complete
ensemble, like the full complement of $d+1$ unbiased bases~\cite{rbksc03}. 
Such sets are useful in a modified protocol in which Eve, instead of Alice,
creates bipartite states and distributes
one half to each of the two parties~\cite{bceekm03}, and as a test to ensure secret key generation 
is possible~\cite{cll03}. Now Alice and Bob need to perform 
quantum state tomography on their pieces to ensure the security of the
protocol. Using the $d^2$ equiangular states again offers more security in this case.


Investigating the intercept/resend attack yields an insight into how the spherical
codes ultimately perform in comparison with unbiased bases, when the strongest
eavesdropping attacks are considered.  
The spherical codes' flexibility in $n$ and $m$ leads to advantages in
both speed and security, though not together in any one instance, all the while
rendering unnecessary the procedure of sacrificing key letters to determine the error rate.
For fixed resources, i.e. fixed number of states in a given dimension, spherical
codes offer higher noise security thresholds, but slower key generation rates. 

This hindrance may be possible to overcome, or at least ameliorate, for
Bob's use of Alice's spherical code is almost certainly not his best choice.
That is, assuming that the secrecy capacity will increase with increasing classical 
capacity of the quantum channel,
Bob's optimal measurement is likely \emph{not} the spherical code used by Alice. 
This is evident in two dimensions for the two spherical codes, which 
we may think of in the Bloch-sphere representation as a regular tetrahedron and
three equally-spaced coplanar vectors. By inverting each vector in the ensemble
the resulting combination of spherical code encoder and inverse spherical code
decoder achieves a higher classical capacity. In the case of the trine this is 
known to be the optimal measurement~\cite{sbjoh99} and conjectured to be for
the tetrahedron~\cite{davies78}. These are in fact the two qubit protocols based on the 
repudiation measurement alluded to earlier. Much the same is true in three dimensions,
at least for the case of six equiangular states: an unenlightening numerical maximization 
produced the result that a unitarily-transformed version of the encoding spherical
code yields a capacity of 0.638,
a roughly 50\% improvement over the nominal
capacity of 0.424 when using the same ensemble for encoding and decoding.
An even greater classical capacity can be achieved by using a
repudiation measurement, each outcome of which excludes two possible signal states.
In this case the capacity jumps to $0.734$. 
Neither of these strategies surpass the capacity generated by using two unbiased
bases, $\log[3]/2\approx 0.792$, but they narrow the gap. 

The author acknowledges helpful input from C.~M.~Caves, A.~J.~Scott,
and K.~K.~Manne. This work was supported in part by Office of Naval
Research Grant No.~N00014-00-1-0578.

\end{document}